\documentclass{iauc}
\usepackage{graphicx}

\title[Darwin/TPFI architecture comparison] 
{Characteristics of proposed 3 and 4 telescope configurations for Darwin and TPF-I}

\author[Kaltenegger, Fridlund] 
{Kaltenegger L., $^1$$^,$$^2$%
  \thanks{Present address:Harvard-Smithsonian Center for Astrophysics, MS20, 60 Garden Street, Cambridge, MA 02138, USA }
\and Fridlund M.$^2$}

\affiliation{$^1$Harvard-Smithsonian Center for Astrophysics,
60 Garden St, Cambridge, MA 02138, USA \break 
email: lkaltenegger@cfa.harvard.edu\\[\affilskip]
$^2$European Space Agency ESTEC, P.O.Box 299, NL-2200AG Noordwijk, The Netherlands \break 
email:Malcolm.Fridlund@esa.int}

\pubyear{2005}
\volume{IAUC200} 
\pagerange{119--126}
\date{Nov 30th}
\jname{Direct Imaging of Exoplanets: Science and Techniques}
\editors{C. Aime, F. Vakili, eds.}
\begin{document}

\maketitle

\begin{abstract}
The Darwin and TPF-I missions are Infrared free flying interferometer missions based on nulling interferometry. Their main 
objective is to detect and characterize other Earth-like planets, analyze the composition of their atmospheres and their 
capability to sustain life, as we know it. Darwin and TPF-I are currently in definition phase. A number of mission 
architectures of 3 and 4 free flying telescopes are evaluated on the basis of the interferometer's response, ability to 
distinguish multiple planet signatures and starlight rejection capabilities. The characteristics of the new configurations 
are compared also to the former, more complex Bowtie baseline architectures as well as evaluated on base of their science 
capability.



\end{abstract}

\firstsection 
\section{Introduction}
The closing years of the 20th century have allowed us, for the first time, to seriously discuss interferometric instruments 
deployed in space achieving unprecedented spatial resolution. And thus the direct detection of Earth-like exoplanets orbiting 
nearby stars, and the search for key tracers of life in their atmospheres, are high-priority objectives in the long-term 
science plan of ESA as well as NASA. Both Agencies are working on the definition of the instruments that will meet this 
challenge, Darwin and Terrestrial Planet Finder (TPF) respectively with a foreseen launch in 2015+. The baseline mission 
duration is 5 years, extendable to 10 years in an L2 orbit. Operating in the infrared band requires that all optical components 
are cooled to roughly 40 K, this is achieved by passive cooling. Only the detector requires active cooling. Darwin is a 
major element in the Cosmic Vision 2020 program of the European Space Agency. It has the explicit purpose of detecting other 
Earth-like worlds, analyze their characteristics, determine the composition of their atmospheres and investigate their 
capability to sustain life as we know it. The Darwin mission is envisioned as four free flying spacecraft including 
one beam-combining spacecraft. The beam combiner and the telescope spacecraft fly in one plane with each telescope spacecraft 
at the same distance from the beam combiner. The resolution of the interferometer is adjusted by changing the distance between 
the telescope spacecrafts. A similar activity has been taking place in the United States within the context of NASA's Origins 
program. A science collaboration has already been established. The missions are currently in definition phase, here some 
characteristics of the designs are reviewed. We present the different new configuration architectures under investigation. 

\section{Nulling Interferometry}
The main problem in the direct detection of an exo-planet of a size comparable to our own Earth and located at a similar 
distance from its own star is one involving contrast and dynamical range. A central star like our Sun (G2V) 
outshines an Earth-like planet in the visual wavelength range by a factor of at least $10^9$. Going to the mid-infrared 
alleviates this problem, because the planet's thermal emission peaks at 10$\mu$m. 
Even at these wavelengths, the contrast is more than a factor of $10^6$. Analysis of the planetary light requires that the 
stellar light is suppressed to a high degree. In the IR this is done by a technique called nulling interferometry, in essence 
this means that achromatic phase shifts are applied to the beams collected by individual telescopes before recombination such 
that the on-axis light, i.e. stellar light, is cancelled by destructive interference, while the much weaker planetary light 
emitted at a certain off-axis angle interferes constructively. By keeping a star in the center of the image plane, coronography 
is realized without the presence of a physical mask. The output of the system can be described by an angular transmission map 
(TM) featuring interference fringes, with a sharp null (destructive interfered area) in the center of the map. The TM gives 
the interferometer's intensity response as a function of the sky coordinates. The modulation map (MM) is the difference between 
the recorded outputs on detector A and detector B. The stellar signal is nulled out only on the optical axis. A leakage of 
photons out of the central null exists because the star has a finite photospheric disk. That leakage is a very important noise 
source. The actual shape and transmission properties of the pattern are a function of the number of telescopes, configuration, 
and the distance between the telescopes (see e.g, \cite[Absil (2001)]{Absil01},\cite[Kaltenegger (2004)]{Kaltenegger04}). Information about the distribution of 
planets in the target star system can be recovered by modulation of that signal. Figure~\ref{fig1} shows the TM, MM and signal modulation for a 
simulated Solar System (Earth, Venus, Mars, Jupiter shown) for different configurations. No spatial information is extracted 
in a single exposure. Rotation modulates the interferometer output intensity as a planet passes in and out of the dark fringes. 
From the intensity and actual pattern of this modulation one can derive the planet's parameters. 

\subsection{Architecture design}
The original Darwin mission concept was optimized for high stellar rejection to focus on observing the closest stars for planetary 
companions. This led to a baseline concept of a free flying configuration with 6 collector telescopes (Bowtie configuration) of 1.5 m diameter and a central beam combiner (see \cite[Absil (2001)]{Absil01}). Optimizing the mission for nearby stars is not the same as optimizing the mission 
performance for its overall star sample of a minimum of 165 stars out of the DARWIN target catalogue \cite[Kaltenegger, Eiroa, Stankov \etal\ (2005)]{Kaltenegger05}. Recent analyses show that the starlight rejection criteria can be relaxed while still maintaining a target sample of 165 stars. The integration time for nearby stars increases, but the overall mission performance does not degrade significantly. A number of these alternative mission architectures have been evaluated on the basis 
of interferometer response, achievable modulation efficiency, number of telescopes and starlight 
rejection capabilities (see \cite[Kaltenegger \& Karlsson (2004)]{KaltKarl04} and \cite[Lay \& Dubovitsky (2004)]{LayDub04}). 
Accordingly, candidate Darwin and TPFI configurations use three (Three telescope Nuller (TTN) see 
\cite[Karlsson, Wallner \& Perdigues Armengol (2003)]{Karlsson03}) or four (X-array, see \cite[Lay \& Dubovitsky (2004)]{LayDub04} telescopes of 3.5 m diameter, reducing complexity 
and cost of the mission. Three telescopes is the minimum number of telescopes needed for an interferometer mission that uses 
rapid signal modulation to detect a planet in the high background noise. An array with 6 telescopes like the Bowtie achieves 
better performance in starlight rejection than an array with 3 or 4 telescopes, but adds complexity and cost to the mission. 
Furthermore the maximum modulation efficiency is 100\%, 93\% while only 70\% for the X-array, TTN and Bowtie, repectively. The mean modulation efficiency is 26\%, 32\% and only 16\% for the X-array, TTN and Bowtie, repectively. These numbers show a 
performance increase for the 3 and 4 telescope architectures in comparison to the Bowtie. Modulation efficiency is a measure of the efficiency with which 
planet photons are converted into output signal. The maximum modulation efficiency sets the performance for the mission's 
spectroscopy phase, the mean modulation the efficiency for the search phase. The higher the modulation efficiency, the better the overall 
performance of the architecture. 

\begin{figure}
\centering
\resizebox{8.2cm}{!}{\includegraphics{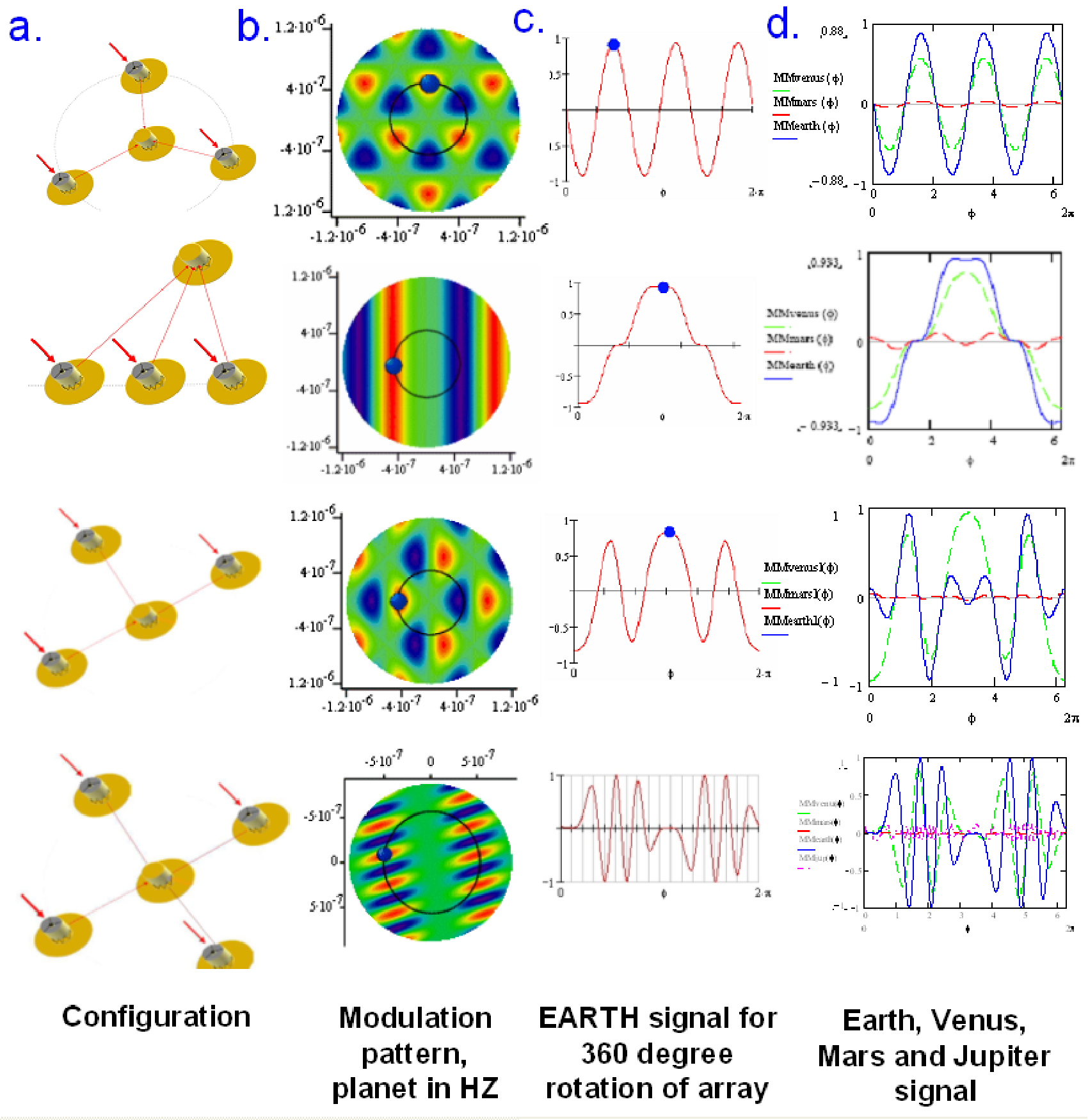} }
\resizebox{4.8cm}{!}{\includegraphics{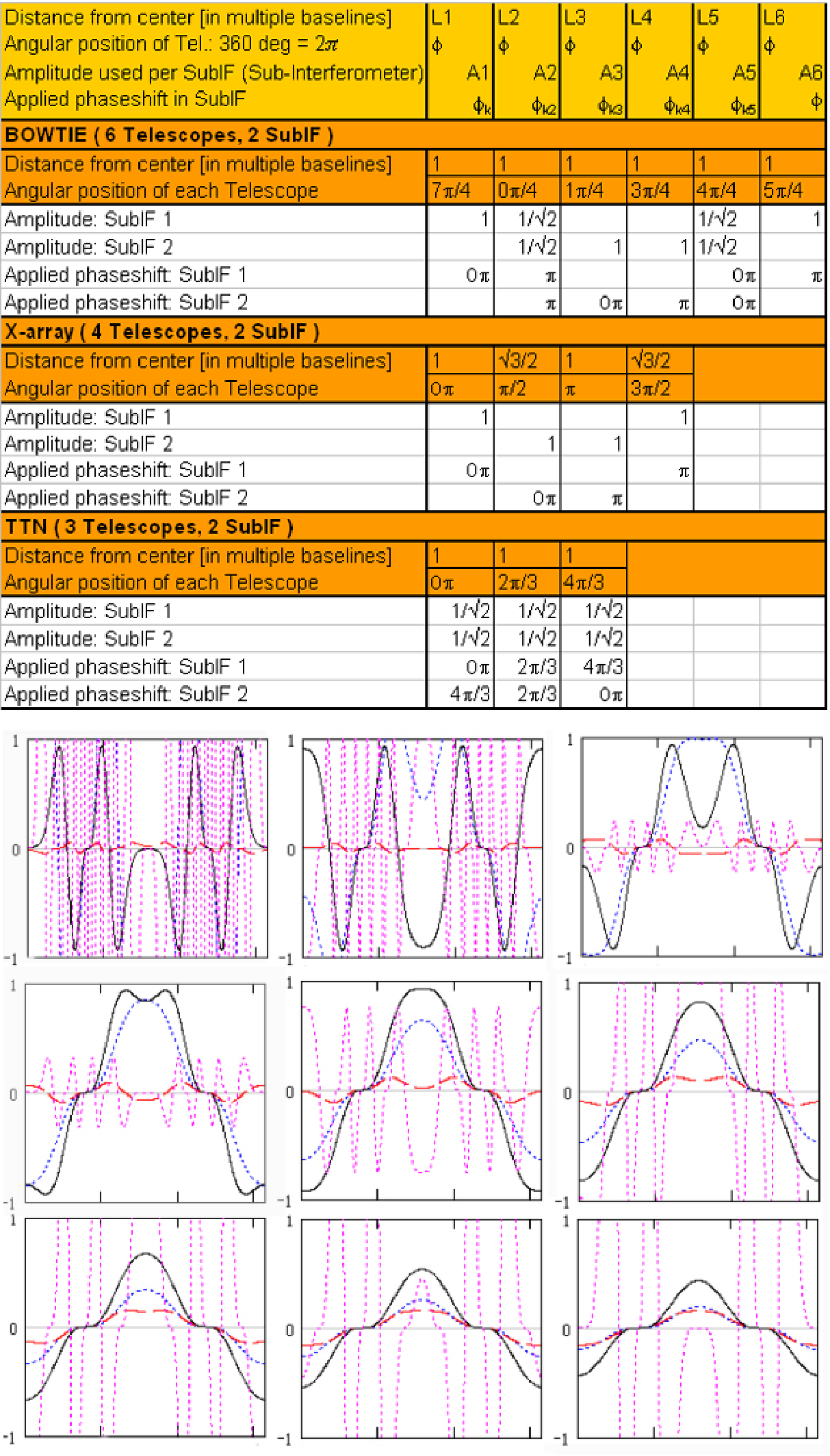} }
  \caption{a) Telescope Array Configuration b) Modulation pattern generated by the array c) modulation of the signal of an 
  Earth-like planet at 1AU over a full 360 degree rotation of the array. d) Modulation of the signal from multiple planets 
  (simulation of our Solar System, planets shown: Venus, Earth, Mars and Jupiter). Different frequency of the signal from 
  different planets allows to disentangle the signal from multiple planets. Configurations shown: upper panel to lower panel: 
  1) equilateral triangle, 2) linear, and 3) rectangular triangle Three Telescope Nuller (TTN) and 4) X array. 
  Right upper panel: Table with architecture characteristics. Right lower panel: The signal of multiple planets in our solar 
  system  (Venus, Earth, Mars and Jupiter) recorded in different spectral channels over a whole rotation of the array for the 
  linear TTN configuration (assuming no noise). The high frequency signal shown is Jupiter's. The panels show the monochromatic signal per channel from 4$\mu$m (left) 
  to 20$\mu$m (right) in steps of 2$\mu$m. }\label{fig1}
\end{figure}

The output beam comprises all information from a specific set of sources on the sky, as well as the background. In an ideal 
scenario with only a single planet around its host star and no other disturbing sources, such as extrasolar zodiacal dust in 
the target system, the detection of a positive flux would imply that a planet is present, if the star is well and truly `nulled 
out'. In real observations, several factors affect the signal to noise in a detrimental way. This has been analyzed by 
different groups e.g. \cite[Lay \& Dubovitsky (2004)]{LayDub04} and \cite[Kaltenegger \& Karlsson (2004)]{KaltKarl04}. Figure~\ref{fig1} shows that 
it is essential to apply spectral channels to distinguish the signal from multiple planets due to their distinct signature as 
a function of distance from their star, wavelength and rotation of the array (shown here, Venus, Earth, Mars and Jupiter). 
The simulations of our own Solar system in Figure~\ref{fig1} show that Jupiter's signal is seen at higher frequency over a 
full rotation of the array than an Earth-like planet. Thus a Jupiter will be easily distinguished. Our simulations show that Venus and Earth are the two planets 
in our own solar system that are most likely to be confused. Here the different architectures show their capability to 
disentangle multiple planetary signals. As seen in Figure~\ref{fig1} the rectangular TTN and the X-array perform the best in 
this respect. Note that these calculations do not take noise into account. By increasing the distance between the telescopes any modulation pattern can be made finer, but that generally increases the 
stellar leakage. The X-array concept has an independent nulling and imaging baseline (the imaging baseline controls the 
resolution of the modulation map). It is the distance between the two sub-nulling Interferometers in the X-array concept. 
Preliminary investigations show that the 4 telescope array (X-array) has a better performance distinguishing multiple planets 
because its modulation pattern has a finer grid \cite[Lay \& Dubovitsky (2004)]{LayDub04}. The TTN design is a simplified concept that combines the nulling and imaging 
baseline. Increasing the imaging baseline to gain resolution thus also increases the nulling baseline what decreases the extent 
of the central null as the whole modulation map scales and thus increases the stellar leakage. 
Recent simulations by \cite[Velusamy \& Marsh (2005)]{Velu05} and \cite[Thiebaut \& Mugnier (2005)]{Thie05} argue that high spatial 
resolution is not needed to disentangle multiple planets, avoiding the difference in performance due to leakage increase for 
the TTN at larger baselines. This leads to a similar performance of the three and four telescope architectures.

\section{Conclusions}\label{sec:concl}
A number of mission architectures of 3 and 4 free-flying telescopes are evaluated on the basis of the interferometer's 
response, modulation efficiency, ability to distinguish multiple planet signatures and starlight rejection capabilities. 
Figure~\ref{fig1} shows the characteristics of the alternative mission architectures. Even though the starlight rejection 
properties of the Bowtie configuration is outstanding in the comparison, the higher mean modulation efficiency of the TTN 
configuration and 
the X-array counteracts that superiority. An additional factor in mission design and complexity is the number of telescopes 
used. Here the Bowtie has a big disadvantage, as it needs 6 telescopes. 
Figure~\ref{fig1} shows that it is essential to apply spectral 
channels to distinguish the signal from multiple planets due to their distinct signature as a function of distance from their 
star, wavelength and rotation of the array (shown here, Venus, Earth, Mars and Jupiter). Here the different architectures 
show their capability to disentangle multiple planetary signals. As seen in Figure~\ref{fig1} the rectangular TTN and the X-array perform 
the best in this respect. Recent simulations by Velusamy et al. and Thiebaut et al (these proceedings) argue that high spatial 
resolution is not needed to disentangle multiple planets which leads to a similar performance of the three and four telescope 
architectures. 


\end{document}